\documentclass[journal]{IEEEtran}
\usepackage{mathrsfs}
\usepackage{amsfonts}
\usepackage{ifpdf}
\usepackage{cite}
\usepackage{stfloats}
\ifCLASSINFOpdf
\usepackage[pdftex]{graphicx}
\else \fi
\usepackage[cmex10]{amsmath}
\usepackage{array}
\usepackage{mdwmath}
\usepackage{mdwtab}
\usepackage{eqparbox}
\usepackage[tight,footnotesize]{subfigure}
\usepackage{algorithmic}
\usepackage{algorithm}
\usepackage{amsmath}
\usepackage{epsfig}
\usepackage{ae}
\usepackage{amssymb}
\usepackage{times}
\usepackage{amsmath}   
\usepackage{booktabs}  
\usepackage{threeparttable} 
\usepackage{multirow}       
\usepackage{xcolor}
\usepackage{paralist}
\newtheorem{theorem}{Theorem}
\newtheorem{lemma}{Lemma}
\newtheorem{corollary}{Corollary}

\newtheorem{remark}{Remark}

\begin{document}

\title{End-to-end Delay Modeling in Buffer-Limited MANETs: A General Theoretical Framework}

\author{Jia Liu$^{1,2}$, Min Sheng$^2$, {\em Member, IEEE}, Yang Xu$^3$, Jiandong Li$^2$, {\em Senior Member, IEEE}, and Xiaohong Jiang$^1$, {\em Senior Member, IEEE}
\thanks{$^1$J.~Liu and X.~Jiang are with the School of Systems Information Science, Future University Hakodate, Kamedanakano 116-2, Hakodate, Hokkaido, 041-8655, Japan. Email: jliu871219@gmail.com, jiang@fun.ac.jp}
\thanks{$^2$J.~Liu, M.~Sheng (corresponding author) and J.~Li are with the State Key Laboratory of ISN, Xidian University, Xi'an, Shaanxi, 710071, China. Email: liujia@mail.xidian.edu.cn, mshengxd@gmail.com, jdli@ieee.org}
\thanks{$^3$Y.~Xu is with the School of Economics and Management, Xidian University, Xian, Shaanxi, 710071, China. Email: yxu@xidian.edu.cn}
} 
\maketitle

\begin{abstract}
This paper focuses on a class of important two-hop relay mobile ad hoc networks (MANETs) with limited-buffer constraint and any mobility model that leads to the uniform distribution of the locations of nodes in steady state, and develops a general theoretical framework for the end-to-end (E2E) delay modeling there. We first combine the theories of Fixed-Point, Quasi-Birth-and-Death process and embedded Markov chain to model the limiting distribution of the occupancy states of a relay buffer, and then apply the absorbing Markov chain theory to characterize the packet delivery process, such that a complete theoretical framework is developed for the E2E delay analysis. With the help of this framework, we derive a general and exact expression for the E2E delay based on the modeling of both packet queuing delay and delivery delay. To demonstrate the application of our framework, case studies are further  provided under two network scenarios with different MAC protocols to show how the E2E delay can be analytically determined for a given network scenario. Finally, we present extensive simulation and numerical results to illustrate the efficiency of our delay analysis as well as the impacts of network parameters on delay performance.

\end{abstract}


\section{Introduction} \label{section:introduction}

With the development of wireless communication technologies, mobile ad hoc networks (MANETs) have become an appealing candidate for many critical applications, such as emergency rescue, disaster relief, coverage extension for cellular networks, etc. \cite{Tanenbaum_BOOK2013,Perkins_BOOK01,Ramanathan_Magzine2002,Andrews_CM08,Goldsmith_CM11}. Although lots of work has been done to facilitate the commercialization of MANETs, understanding their fundamental delay performance has been a critical research issue for them to support various applications with different quality of service (QoS) requirements \cite{Hanzo_Survey2007,Chen_Survey2007}. 

End-to-end (E2E) delay, the time that a packet takes to reach its destination after it is generated by its source, serves as the most fundamental delay metric. The available theoretical studies on E2E delay of MANETs mainly focus on deriving its upper bound or approximation. Regarding the delay upper bound of MANETs, Neely \emph{et al.} \cite{Neely_IT05} derived some useful results for a cell-partitioned MANET with the two-hop relay (2HR) routing scheme and i.i.d mobility model. Later, Gamal \emph{et al.} \cite{Gamal_IT06} and Sharma \emph{et al.} \cite{Sharma_TON07} extended the results of \cite{Neely_IT05} to the continuous network model and general mobility model, respectively. Inspired by these works, extensive research activities have been devoted to the study of delay upper bound for MANETs under various network scenarios, such as under the motioncast in \cite{Wang_TON11}, under the cognitive networks in \cite{Wang_INFOCOM11}, under the packet redundancy in \cite{Liu_TWC11}, under the multi-hop back-pressure routing in \cite{Alresaini_INFOCOM12}, and under the power control in \cite{Gao_IEICE13}. Regarding the delay approximation, Jindal \emph{et al.} \cite{Jindal_TMC2009} explored recently the E2E delay approximation for MANETs with multi-hop relay routing, and Liu \emph{et al.} studied the E2E delay approximation for MANETs with probing-based 2HR routing \cite{Liu_TWC12} and limited packet redundancy \cite{Liu_WCNC12}.

In addition to the studies on delay upper bound or approximation for MANETs, Neely \emph{et al.} \cite{Neely_IT05} also applied the queueing theory to derive the exact expression for E2E delay. Following this line, recently some results have been reported on the modeling of really achievable E2E delay in MANETs. Chen \emph{et al.} \cite{Chen_ICCC13} explored the MANETs with Aloha MAC protocol and determined the corresponding exact E2E delay there under the continuous network model. For a cell-partitioned MANET with broadcast-based routing scheme, Gao \emph{et al.} \cite{Gao_Wiopt13} proposed a new theoretical framework for the analysis of its exact E2E delay based on the theory of Quasi-Birth-and-Death process.

It is notable that the common limitation of above studies is that to simplify their analysis of E2E delay, they assume the relay buffer of a node, which is used for temporarily storing packets of other nodes, has an infinite buffer size. In a practical MANET, however, the buffer size of a mobile node is usually limited due to both its storage space limitation and computing capability limitation. Thus, for the practical delay performance study of MANETs, the constraint on buffer space should be carefully addressed. Notice that the E2E delay modeling with practical limited-buffer constraint still remains a technical challenge. This is mainly due to the lack of a general theoretical framework to efficiently characterize the highly dynamic behaviors in such networks, like the complicated buffer occupancy states of a relay buffer, as well as the highly dynamic queuing process and delivery process of a packet.

As a step towards the modeling of real achievable E2E delay for the practical MANETs with buffer constraint, we focus on a class of important 2HR MANETs with limited shared relay buffer and propose a general theoretical framework for the E2E delay modeling there. The main contributions of this paper are summarized as follows.

\begin{itemize}

\item
For the concerned MANET, we first combine the theories of Fixed-Point (FP), Quasi-Birth-and-Death (QBD) process and embedded Markov chain (EMC) to construct an analytical model to fully depict the complicated occupancy behaviors of a relay buffer with limited buffer size.

\item
Based on the above modeling of relay buffer occupancy behaviors, we then apply the absorbing Markov chain (AMC) theory to characterize the packet delivery process, such that a complete theoretical framework is developed for the E2E delay modeling in the concerned buffer-limited MANETs. This framework is general in the sense that it can be applied to conduct E2E delay analysis for a 2HR MANET with any mobility model that leads to the uniform distribution of the locations of nodes, such as the i.i.d mobility model \cite{Neely_IT05}, the random walk model \cite{Gamal_IT06}, the random way-point model \cite{Zhou_INFOCOM10}, etc..

\item 
To demonstrate the application of the proposed framework, case studies are further provided under two network scenarios, i.e., the cell partitioned networks with local scheduling-based MAC protocol (LS-MAC) \cite{Neely_IT05} and Equivalent-Class based MAC protocol (EC-MAC) \cite{Kulkarni_IT04}, to show how the E2E delay can be analytically determined for a given network scenario by applying our framework. Finally, extensive simulation and numerical results are provided to validate the efficiency of the proposed E2E delay model and also to illustrate the impacts of network parameters on delay performance.

\end{itemize}

The remainder of this paper is organized as follows. Section~\ref{section:preliminaries} introduces preliminaries involved in this paper. The complicated relay buffer occupancy behaviors are analyzed in Section~\ref{section:relay_buffer_analysis}. We derive the queuing delay, delivery delay and E2E delay in Section~\ref{section:delay}, and conduct case studies in Section~\ref{section:case_studies}. The simulation results and corresponding discussions are provided in Section~\ref{section:simulation}. Finally, Section~\ref{section:conclusion} concludes this paper.

\section{Preliminaries} \label{section:preliminaries}

In this section, we first present some basic assumptions and the buffer constraint, and then introduce the routing scheme and some critical definitions involved in this study.

\subsection{Basic Assumptions} \label{subsection:assumption}
We consider the following minimal set of assumptions:

\begin{enumerate}[({A}.i)] 

\item
The ad hoc network is time-slotted and consists of $n$ mobile nodes. 

\item
The packet generating process in each source node is independent and assumed to be a Bernoulli process, where a packet is generated by its source node with probability $\lambda$ in a time slot. 

\item
The widely-used permutation traffic model \cite{Neely_IT05,Grossglauser_TON02,Ciullo_TON11} is adopted. With this traffic model, there are $n$ unicast traffic flows in the network, each node is the source of one traffic flow and also the destination of another traffic flow. We denote by $\varphi(i)$ the destination node of the traffic flow originated from node $i$, then the source-destination pairs are matched in a way that the sequence $\{\varphi(1),\varphi(2),\cdots,\varphi(n)\}$ is just a derangement of the set of nodes $\{1,2,\cdots,n\}$.

\item
During a time slot the total amount of data that can be transmitted from a transmitter to its corresponding receiver is fixed and normalized to one packet.

\item
We consider the mobility model that leads to the uniform distribution of the locations of nodes in steady state, which covers many typical mobility models such as the i.i.d mobility model, the random walk model, the random way-point model, etc.. More formally, we denote by $X_i(t)$ the location of $i$th node at time slot $t$ and assume the process $\left\{X_i(\cdot) \right\}$ is stationary and ergodic with stationary distribution uniform on the network area; moreover, the trajectories of different nodes are independent and identically distributed. 

\end{enumerate}

\subsection{Buffer Constraint}

\begin{figure}[!t]
\centering
\includegraphics[width=3.5in]{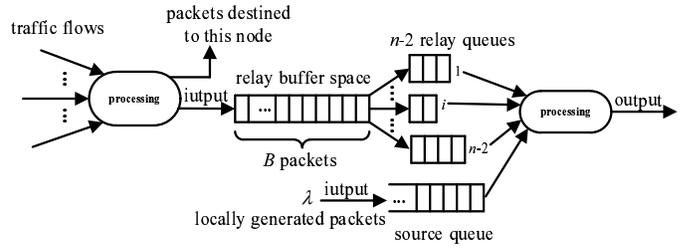} \caption{Illustration of buffer structure of a node.}
\label{fig:buffer_constraint}
\end{figure}

As illustrated in Fig.~\ref{fig:buffer_constraint}, each node in the MANET maintains $n-1$ individual queues, one source queue for storing the packets that are locally generated at this node, and $n-2$ parallel relay queues for storing packets of other flows (one queue per flow). All these queues follow the FIFO (first-in-first-out) discipline.

Similar to the available studies on buffer-limited wireless networks \cite{Le_TON12,Herdtner_INFOCOM05}, we consider the following practical buffer constraint that all the $n-2$ relay queues of a node share a common relay buffer  with the limited buffer size of $B$ packets, while the buffer size of source queue is unlimited. We adopt this buffer constraint here mainly due to the following reasons. First, the mathematical tractability of this assumption allows us to gain important insights into the structure of E2E delay analysis. Second, the analysis under this assumption provides a meaningful theoretical result in the limit of infinite source buffer. Third, in a practical wireless network, each node usually prefers to reserve a much larger buffer space for storing its own packets than that for storing packets of other flows. Also, even in the case that the buffer space of source queue is not enough when bursty traffic comes, the congestion control in the upper layer can be executed to avoid the loss of locally generated packets \cite{Le_TON12}.

\subsection{Handshake-based 2HR Scheme}
Regarding the routing scheme, we focus on the 2HR scheme, because it is simple yet efficient and thus serves as a class of attractive routing protocols for MANETs \cite{Grossglauser_TON02,Neely_IT05}. To avoid unnecessary packet loss and support the efficient operation of the concerned buffer-limited MANETs, we introduce a handshake mechanism with negligible overhead\footnote{The handshake mechanism can be easily implemented by sending only one indicator bit from the receiver to the transmitter (e.g., bit 0 when the relay buffer is full, and bit 1 otherwise), so the impact of this overhead can be neglected in our analysis.} into the 2HR scheme such that the packet dropping will not happen even in the case of relay buffer overflow. Once a node (say $\mathbf{S}$) gets access to the wireless channel in a time slot, it executes the new handshake-based 2HR (H2HR for short) routing scheme summarized in Algorithm~\ref{algorithm:H2HR}. 

\begin{algorithm}[!ht]
\caption{H2HR algorithm}
\label{algorithm:H2HR}
\begin{algorithmic}[1]
\IF{The destination $\mathbf{D}$ is within the transmission range of $\mathbf{S}$}
  \STATE $\mathbf{S}$ executes \textbf{Procedure}~\ref{procedure:s-d}.
\ELSIF{There exist other nodes within the transmission range of $\mathbf{S}$}
	\STATE With equal probability, $\mathbf{S}$ selects one node as the receiver.
  \STATE $\mathbf{S}$ executes \textbf{Procedure}~\ref{procedure:s-r} or \textbf{Procedure}~\ref{procedure:r-d} equally with the receiver.	
\ENDIF
\end{algorithmic}
\end{algorithm}

\renewcommand\thealgorithm{1}
\floatname{algorithm}{Procedure}
\begin{algorithm}[!ht]
\caption{Source-to-destination (\textbf{S-D}) transmission}
\label{procedure:s-d}
\begin{algorithmic}[1]
\IF{$\mathbf{S}$ has packets in its source queue}
  \STATE $\mathbf{S}$ transmits the head-of-line (HoL) packet in its source queue to $\mathbf{D}$.
	\STATE $\mathbf{S}$ removes the HoL packet from its source queue. 
	\STATE $\mathbf{S}$ moves ahead the remaining packets in its source queue.
\ELSE
  \STATE $\mathbf{S}$ remains idle.
\ENDIF
\end{algorithmic}
\end{algorithm}

\renewcommand\thealgorithm{2}
\begin{algorithm}[!ht]
\caption{Source-to-relay (\textbf{S-R}) transmission}
\label{procedure:s-r}
\begin{algorithmic}[1]
\IF{$\mathbf{S}$ has packets in its source queue}
  \STATE $\mathbf{S}$ \textbf{initiates a handshake with the receiver to check whether the relay buffer of receiver is full or not.}
	\IF{The relay buffer of receiver does not overflow}
	  \STATE The receiver dynamically allocates a new buffer space to the end of the corresponding relay queue.
    \STATE $\mathbf{S}$ transmits the HoL packet in its source queue to the receiver.
	  \STATE $\mathbf{S}$ removes the HoL packet from its source queue. 
		\STATE $\mathbf{S}$ moves ahead the remaining packets in its source queue.
	\ENDIF	
\ELSE
  \STATE $\mathbf{S}$ remains idle.
\ENDIF
\end{algorithmic}
\end{algorithm}

\renewcommand\thealgorithm{3}
\begin{algorithm}[!ht]
\caption{Relay-to-destination (\textbf{R-D}) transmission}
\label{procedure:r-d}
\begin{algorithmic}[1]
\IF{$\mathbf{S}$ has packets destined to the receiver}
  \STATE $\mathbf{S}$ transmits the HoL packet in its corresponding relay queue to the receiver.
	\STATE $\mathbf{S}$ removes the HoL packet from this relay queue. 
	\STATE $\mathbf{S}$ moves ahead the remaining packets in this relay queue.	
	\STATE This relay queue releases one buffer space to the common relay buffer of $\mathbf{S}$.
\ELSE
  \STATE $\mathbf{S}$ remains idle.
\ENDIF
\end{algorithmic}
\end{algorithm}

\subsection{Definitions}

Here we introduce some important definitions involved in this study.

\textbf{Relay-buffer Overflowing Probability (ROP)}: For the concerned MANET with a given packet generating rate $\lambda$ in each node, the relay-buffer overflowing probability $p_o(\lambda)$ of a node is defined as the probability that the relay buffer of this node overflows (i.e, the relay buffer is full).

\textbf{Queuing Delay}: The queuing delay is defined as the time it takes a packet to move to HoL in the source queue (i.e., the source node starts to deliver it) after it is generated by its source.

\textbf{Delivery Delay}: The delivery delay is defined as the time it takes a packet to reach its destination after its source starts to deliver it.

\textbf{End-to-end Delay}: The end-to-end delay is defined as the time it takes a packet to reach its destination after it is generated by its source, which is the sum of its queuing delay and delivery delay.

\section{Relay Buffer Analysis} \label{section:relay_buffer_analysis}
In this section, we first introduce three basic probabilities. Based on these probabilities, we then apply the QBD process modeling and EMC technique to depict the occupancy behaviors of a relay buffer. Finally, we construct a self-mapping function for the ROP $p_o(\lambda)$ (i.e., $p_o(\lambda)$ is the fixed-point of this function) to determine the limiting distribution of the occupancy states, which will help us conduct delay analysis in Section~\ref{section:delay}.

Due to the symmetry of nodes and traffic flows, we only focus on one node $\mathbf{S}$ in the following analysis. We denote by $p_{sd}$, $p_{sr}$ and $p_{rd}$ the probabilities that in a time slot $\mathbf{S}$ gets access to the wireless channel and decides to execute \textbf{S-D}, \textbf{S-R} and \textbf{R-D} transmission respectively\footnote{It is notable that $p_{sr}=p_{rd}$, and executing a transmission doesn't mean that $\mathbf{S}$ will successfully transmit a packet in this time slot.}. These probabilities can be determined under a given network scenario and the derivation of them will be elaborated in case studies.

\subsection{QBD Process Modeling}
Regarding the source queue of $\mathbf{S}$, it can be modeled as a Bernoulli/Bernoulli queue \cite{Daduna_BOOK01} with packet arrival rate $\lambda$ and service rate $\mu_s(\lambda)$, where $\mu_s(\lambda)$ is given by
\begin{equation} \label{eq:mu_s}
\mu_s(\lambda)=p_{sd}+p_{sr}\left(1-p_o(\lambda) \right).
\end{equation}
Due to the reversibility of Bernoulli/Bernoulli queue, the packet departure process of source queue is also a Bernoulli process with rate $\lambda$.

Regarding the relay buffer of $\mathbf{S}$, we adopt a two-tuple $\mathbf{X}(t)=\left(I(t),J(t)\right)$ to define its state at time slot $t$, where $I(t)$ denotes the number of packets occupying the relay buffer, and $J(t)$ denotes the number of relay queues which are not empty, here $0 \leq I(t) \leq B$, $1 \leq J(t) \leq I(t)$ when $I(t)>0$, and $J(t)=0$ when $I(t)=0$.

\begin{figure}[!t]
\centering
\includegraphics[width=2.5in]{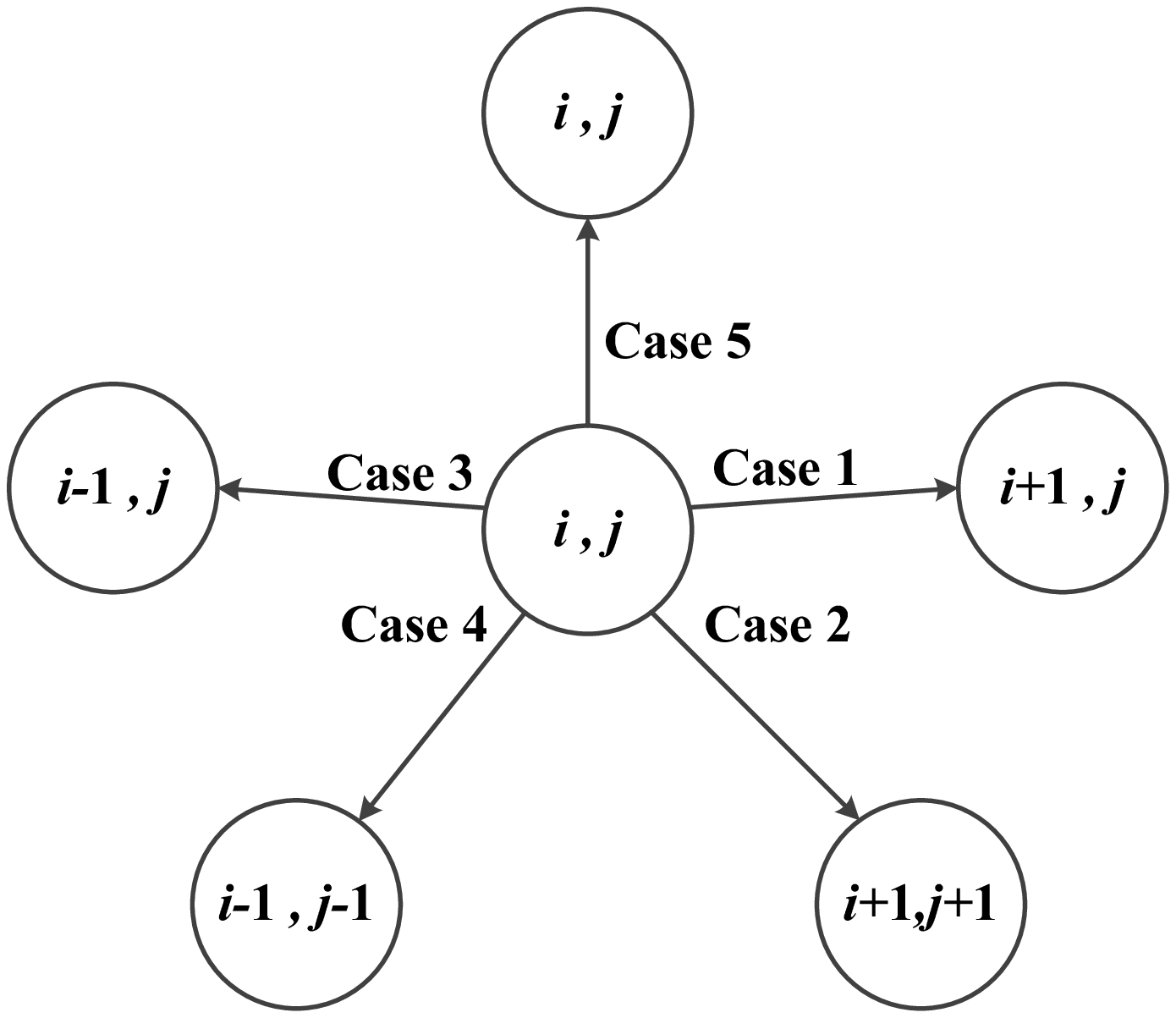} \caption{Transition cases from a general state $(i,j)$.}
\label{fig:transition_cases}
\end{figure}

As illustrated in Fig.~\ref{fig:transition_cases}, suppose that the relay buffer of $\mathbf{S}$ is in state $(i,j)$ at the current time slot, only one of the following transitions may happen in the next time slot:

\begin{enumerate}[{Case} 1:]

\item
$i<B$, a packet enters the relay buffer, and this packet is destined for a destination same as one of packet(s) already in relay queues. 

\item
$i<B$, a packet enters the relay buffer, and the destination of this packet is different from all packet(s) already in relay queues. 

\item
$i>0$, a packet in one of the relay queues is delivered to its destination, and there still exist other packet(s) in this relay queue. 

\item
$i>0$, a packet in one of the relay queues is delivered to its destination, and there is no remaining packet in this relay. 

\item
no packet enters into or departs from the relay queues. 

\end{enumerate}

\begin{figure}
\centering
\includegraphics[width=3.0in]{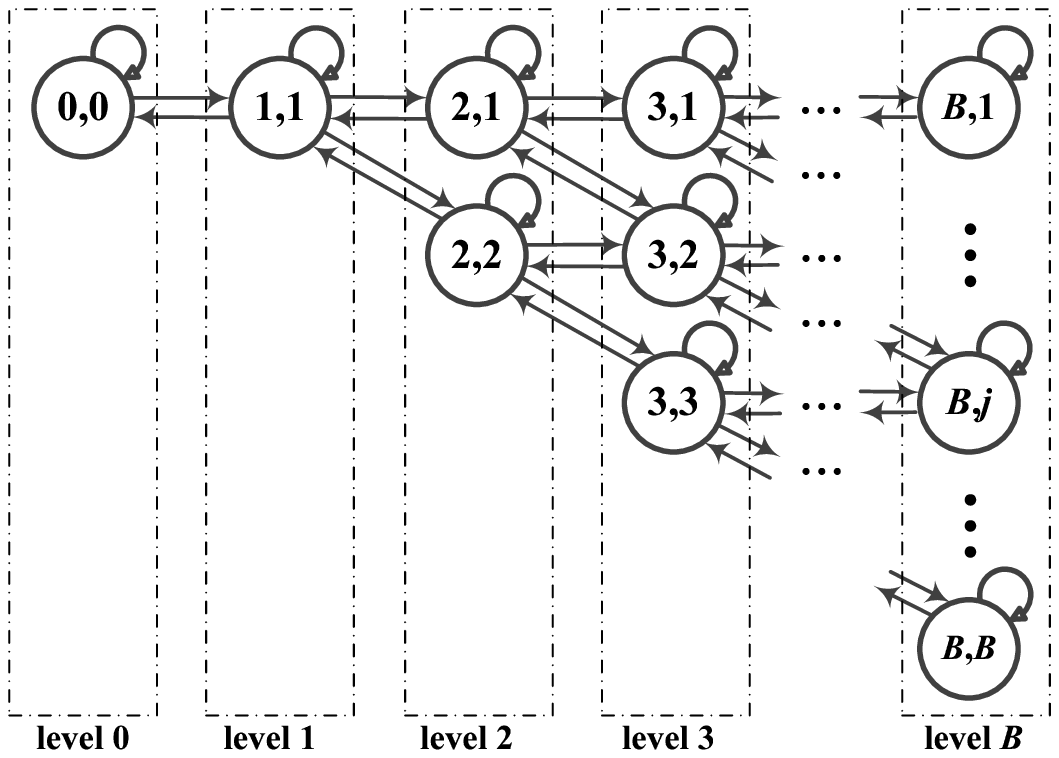}
\caption{State transition diagram of the QBD process.}
\label{fig:state_machine}
\end{figure}

To facilitate our discussion, we call the subset of states $L_i=\left \{(i,1),(i,2),\cdots,(i,i) \right \}$ level $i$, $L_0=\left \{(0,0) \right \}$ level $0$, and state $(i,j)$ that the relay buffer is in level $i$ and phase $j$. Notice that when the relay buffer is in some state of level $i$ at a time slot, the next state after one-step state transitions could only be some state in the same level or its adjacent levels. Thus, as time evolves, the state transitions of the relay buffer of $\mathbf{S}$ form a two-dimensional QBD process $\left \{ \mathbf{X}(t), t=0,1,2,\cdots \right \}$ \cite{Alfa_BOOK10,Latouche_BOOK99}. According to the transition cases in Fig.~\ref{fig:transition_cases}, the overall transition diagram of the QBD process is summarized in Fig.~\ref{fig:state_machine}. There are in total $1+0.5B(1+B)$ states for the QBD process, and we arrange all these states in a low-to-high level and low-to-high phase way as follows: $\left \{(0,0),(1,1),(2,1),(2,2),\cdots,(B,B)  \right \}$. Then the corresponding state transition matrix $\mathbf{P}$ of the QBD process can be determined as
\begin{equation}      
\mathbf{P}\!=\!\left[                 
\begin{array}{ccccc}   
\! \mathbf{A_{0,0}} \!  &  \! \mathbf{A_{0,1}} \!  & \!                              \!  &  \!                               \!  &  \!                        \!  \\  
\! \mathbf{A_{0,1}} \!  &  \! \mathbf{A_{1,1}} \!  & \! \mathbf{A_{1,2}}             \!  &  \!                               \!  &  \!                        \!  \\  
\!                  \!  &  \! \ddots           \!  & \! \ddots                       \!  &  \!  \ddots                       \!  &  \!                        \!  \\
\!                  \!  &  \!                  \!  & \! \mathbf{A_{B\!-\!1,B\!-\!2}} \!  &  \!  \mathbf{A_{B\!-\!1,B\!-\!1}} \!  &  \! \mathbf{A_{B\!-\!1,B}} \! \\
\!                  \!  &  \!                  \!  & \!                              \!  &  \!  \mathbf{A_{B,B\!-\!1}}       \!  &  \! \mathbf{A_{B,B}}       \!
\end{array}
\right],   \label{eq:QBD_matrix}             
\end{equation}
where the sub-matrix $\mathbf{A_{i,l}}$ is of size $i\times l$ ($\mathbf{A_{0,0}}$, $\mathbf{A_{0,1}}$ and $\mathbf{A_{1,0}}$ are of size $1\times 1$), denoting the transition probabilities from the states of level $i$ to the states of level $l$.

It is notable that in our QBD process of relay buffer, different levels have different number of phases, and the transition probabilities of one state depend on its level, thus the QBD process is level-dependent and it is very difficult to solve its limiting distribution by determining its critical matrices and conducting recursive algorithm \cite{Alfa_BOOK10,Latouche_BOOK99}. To address this issue, we adopt a Markov chain-collapsing technique \cite {Hachigian_1963,Subramanian_ISIT09} to convert the two-dimensional QBD process to a one-dimensional EMC in the next subsection.

\subsection{Collapsing to an EMC}

\begin{figure}
\centering
\includegraphics[width=3.0in]{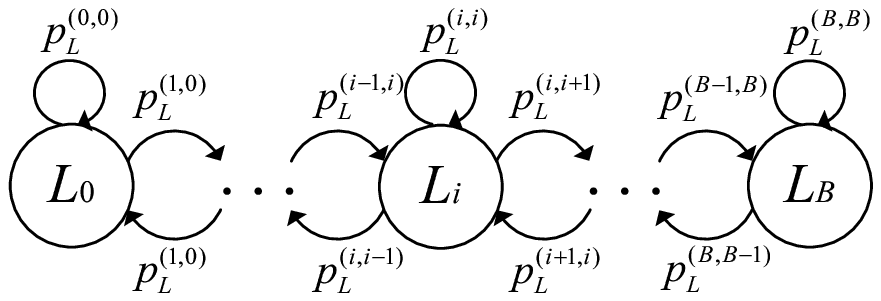} \caption{State machine of the EMC.}
\label{fig:EMC}
\end{figure}
 
For the QBD process of Fig.~\ref{fig:state_machine}, we integrate all states of a level into only one state, then the two-dimensional QBD process is collapsed to a one-dimensional Embedded Markov Chain (EMC). As illustrated in Fig.~\ref{fig:EMC}, one state $L_i$ of the EMC corresponds to one level $i$ of the QBD process, and $p_L^{(i,l)}$ denotes the one-step transition probability from state $L_i$ to state $L_l$ in the EMC. According to the EMC theory \cite{Hachigian_1963,Subramanian_ISIT09}, the state transition probability of the EMC is the phase-averaged state transition probability of the QBD process, then we have
\begin{equation}
p_L^{(i,l)}=\left\{
\begin{aligned}
& p_{(0,0),L_l}, & & i=0 \\
& \sum_{j=1}^{i}{p_{(i,j),L_l}}\cdot P_{j|L_i}, & & 1 \leq i \leq B
\end{aligned}
\right.
\label{eq:phase_average}
\end{equation} 
where $p_{(i,j),L_l}$ denotes the transition probability from state $(i,j)$ to the states of level $l$, and $P_{j|L_i}$ denotes the conditional probability that the relay buffer is in phase $j$ given that it is in level $i$. Based on formula~(\ref{eq:phase_average}) as well as the ergodic and uniform features of the distribution of node location, we have the following lemma regarding the transition probabilities of the EMC. 

\begin{lemma} \label{lemma:EMC_transition}
The one-step transition probability $p_L^{(i,l)}$ of the EMC is determined as
\begin{equation}
p_L^{(i,l)}=\left\{ 
\begin{aligned}
&\rho_s(\lambda) \cdot p_{sr},  &   &l=i+1 \leq B \\
&\frac{i}{n-3+i} \cdot p_{rd},  &   &l=i-1 \geq 0 \\
&1-p_L^{(i,i+1)}-p_L^{(i,i-1)}, &   &l=i          \\
&0,                             &   &\text{others} 
\end{aligned}
\right.
\label{eq:EMC_transition}
\end{equation}
where $\displaystyle \rho_s(\lambda)=\frac{\lambda}{\mu_s(\lambda)}=\frac{\lambda}{p_{sd}+p_{sr}(1-p_o(\lambda))}$.
\end{lemma}

\begin{IEEEproof}
See Appendix~\ref{appendix:EMC_transition} for the proof.
\end{IEEEproof}

We arrange all the states of EMC in a low-to-high level way as follows: $\left \{L_0,L_1,\cdots,L_B\right \}$. Then the corresponding state transition matrix $\mathbf{P}_{EMC}$ can be determined as
\begin{equation}       
\mathbf{P}_{EMC}=\left[                 
\begin{array}{cccc}   
p_L^{(0,0)} & p_L^{(0,1)} &                   &                                               \\  
p_L^{(1,0)} & p_L^{(1,1)} & p_L^{(1,2)}       &                                               \\  
            & \ddots      & \ddots            & \ddots                                        \\
            &             & p_L^{(B,B-1)}     & p_L^{(B,B)}
\end{array}
\right].    \label{eq:EMC_matrix}             
\end{equation}

\subsection{Constructing the self-mapping function}

With the help of transition matrix $\mathbf{P}_{EMC}$, we then construct a self-mapping function for $p_o(\lambda)$, i.e., $p_o(\lambda)$ is the fixed-point of this function \cite{Granas_BOOK03}, such that $p_o(\lambda)$ as well as the limiting distribution of the occupancy states of a relay buffer can be determined. 

From the the state machine of EMC in Fig.~\ref{fig:EMC} and the transition matrix $\mathbf{P}_{EMC}$, we can see that: 1) the EMC is irreducible; 2) each state $L_i$ is recurrent; 3) the period of each state $L_i$ is $1$, so each state is aperiodic. Based on these properties, we can conclude that the EMC is ergodic, thus its limiting distribution $\mathbf{\Pi_L}=\left [ \pi_{L_0},\pi_{L_1},\cdots,\pi_{L_B}\right ]$ exists and is unique, and is same as its stationary distribution \cite{Haggstrom_BOOK02}. Then we have
\begin{align}
& \mathbf{\Pi_L} \cdot \mathbf{P}_{EMC} =\mathbf{\Pi_L}, \label{eq:balance_equation} \\
& \mathbf{\Pi_L} \cdot \mathbf{1} = 1, \label{eq:normalization_equation}
\end{align}
where $\mathbf{1}$ is a column vector of size $(B+1) \times 1$ with all elements being $1$, and equation (\ref{eq:normalization_equation}) follows from the normalization property of a probability vector. Combining (\ref{eq:balance_equation}) with (\ref{eq:normalization_equation}) we have
\begin{equation}
\pi_{L_i}=\frac{\mathrm{C}_i \cdot \rho_s(\lambda)^i} {\sum_{k=0}^B{\mathrm{C}_k  \cdot \rho_s(\lambda)^k}},
\label{eq:limiting_distribution_EMC}
\end{equation}
where $\displaystyle \mathrm{C}_i=\binom{n-3+i}{i}$. 

It is notable that the relay buffer overflows when it is in level $B$, then the critical self-mapping function for $p_o(\lambda)$ is constructed as
\begin{equation}
p_o(\lambda)=f\left( p_o(\lambda) \right)=\pi_{L_B}=\frac{\mathrm{C}_B \cdot \rho_s(\lambda)^B} {\sum_{k=0}^B{\mathrm{C}_i  \cdot \rho_s(\lambda)^k}}. \label{eq:ROP}
\end{equation}

Given a packet generating rate $\lambda$, the self-mapping function doesn't contain any unknown parameters except $p_o(\lambda)$. Thus by solving equation (\ref{eq:ROP}), we can determine the ROP $p_o(\lambda)$ corresponding to a given $\lambda$, and the limiting distribution of the EMC can be recursively determined as
\begin{equation}
\pi_{L_i}=p_o(\lambda) \cdot \rho_s(\lambda)^{i-B} \cdot \frac{\mathrm{C}_i}{\mathrm{C}_B}.
\label{eq:limiting_distribution_EMC_recursive}
\end{equation}
The limiting distribution  $\Pi=\left[\pi_{0,0},\pi_{1,1},\cdots,\pi_{i,j},\cdots,\pi_{B,B} \right]$ of the QBD process can be further determined as
\begin{equation}
\pi_{i,j}=\pi_{L_i}\cdot P_{j|L_i}, \label{eq:limiting_distribution_QBD}
\end{equation}
where $\displaystyle P_{j|L_i}$ is given by formula (\ref{eq:conditional_pro}) in Appendix~\ref{appendix:EMC_transition}.

\begin{remark}
Notice that if we don't apply the handshake mechanism, we can also develop the corresponding theoretical framework in the same way to model the relay buffer occupancy process, where the ROP $p_o(\lambda)$ derived in (\ref{eq:ROP}) just corresponds to the packet dropping probability.
\end{remark}

\section{Delay Analysis} \label{section:delay}
With the help of ROP and limiting distribution of occupancy states of a relay buffer, in this section we analyze the delay performance for the concerned buffer-limited MANET. We denote by $Q$, $D$ and $T$ the queuing delay, delivery delay and E2E delay of a packet respectively. The E2E delay of a packet will be derived by computing its queuing delay and delivery delay respectively. The queuing delay will be obtained by analyzing the queuing process of the source queue, while the delivery delay will be derived by modeling the packet delivery process as an AMC and analyzing the time the chain takes to enter the absorbing state.

Before presenting our main results on the delay performance, we first provide the following lemma regarding the per node throughput capacity, which is the maximal packet generating rate the MANET can stably support, and the corresponding delay can then be determined.

\begin{lemma} \label{lemma:throughput_capacity}
For the considered MANET, its per node throughput capacity $\mu$ is given by
\begin{equation}
\mu=p_{sd}+p_{sr}\frac{B}{n-2+B}. \label{eq:throughput_capacity}
\end{equation}

\end{lemma}

\begin{IEEEproof}
See Appendix~\ref{appendix:throughput_capacity} for the proof.
\end{IEEEproof}

\subsection{Queuing Delay}
Considering a given packet generating rate $\lambda$ ($\lambda<\mu$), the corresponding ROP $p_o(\lambda)$ can be obtained by solving equation (\ref{eq:ROP}), and the service rate of source queue $\mu_s(\lambda)$ can be further determined by formula (\ref{eq:mu_s}). Thus, in the following analysis, we use $p_o$ and $\mu_s$ to represent $p_o(\lambda)$ and $\mu_s(\lambda)$ respectively if there is no ambiguous. Notice that the source queue is a Bernoulli/Bernoulli queue, thus its average queue length $\overline{L}_{source}$ is given by \cite{Daduna_BOOK01}
\begin{equation}
\overline{L}_{source}=\frac{\lambda-\lambda^{2}}{\mu_s-\lambda}. \label{eq:L_s}
\end{equation}  
According to the Little's Law \cite{Robertazzi_BOOK12}, the average delay of a packet in its source queue $\mathbb{E}\{D_s\}$ is given by
\begin{equation}
\mathbb{E}\{D_s\}=\frac{1-\lambda}{\mu_s-\lambda}. \label{eq:D_s}
\end{equation} 
Then, the expected queuing delay $\mathbb{E}\{Q\}$ is determined as
\begin{equation}
\mathbb{E}\{Q\}=\mathbb{E}\{D_s\}-\frac{1}{\mu_s}=\frac{\lambda(1-\mu_s)}{\mu_s(\mu_s-\lambda)} \label{eq:queuing_delay}.
\end{equation}

\subsection{Delivery Delay and End-to-end Delay}

We present the following theorem regarding the expected E2E delay of the concerned buffer-limited MANET.

\begin{theorem} \label{theorem:delay}
(\textbf{Main result}) For the concerned MANET with number of nodes $n$, relay buffer size $B$ and packet generating rate $\lambda$ ($\lambda<\mu$), the expected delivery delay $\mathbb{E}\{D\}$ and the expected E2E delay $\mathbb{E}\{T\}$ of a packet are determined as
\begin{align}
& \mathbb{E}\{D\}=\frac{1+(n-2+\Psi_{n,B,\lambda})(1-p_o)}{\mu_s},  \label{eq:delivery_delay} \\
& \mathbb{E}\{T\}=\frac{1-\lambda}{\mu_s-\lambda}+\frac{(n-2+\Psi_{n,B,\lambda})(1-p_o)}{\mu_s}, 
\label{eq:end-to-end_delay}
\end{align}
where $\displaystyle \Psi_{n,B,\lambda}=\frac{\sum_{i=0}^{B-1}{ i \mathrm{C}_i  \cdot \rho_s^i}}{\sum_{i=0}^{B-1}{\mathrm{C}_i  \cdot \rho_s^i}}$.

\end{theorem}

\begin{IEEEproof}
We focus on a packet $y$ which is the HoL packet of the source queue at time slot $t$, then in the next time slot, $y$ will be delivered to its destination with probability $p_{sd}$, be forwarded to a relay node with probability $p_{sr}\cdot (1-p_o)$, and still stay in the source queue with probability $1-\mu_s$. Thus, the delivery process of packet $y$ can be modeled as an absorbing Markov chain as illustrated in Fig.~\ref{fig:absorbing_Markov}, where $S$, $R$ and $D$ denote the states that $y$ is in source queue, forwarded to a relay, and delivered to its destination, respectively. We denote by $\overline{X}_S$ and $\overline{X}_R$ the average transition times from the transient states $S$ and $R$ to the absorbing state $D$, respectively. Then we have
\begin{align}
& \overline{X}_S=1+\overline{X}_S \cdot (1-\mu_s) + \overline{X}_R \cdot p_{sr} (1-p_o),  \\
& \overline{X}_S= \frac{1+\overline{X}_R \cdot p_{sr}(1-p_o)}{\mu_s}. \label{eq:X_S} 
\end{align}

\begin{figure}[!t]
\centering
\includegraphics[width=3.0in]{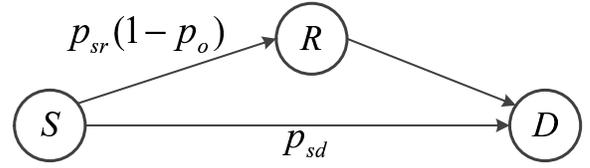}\caption{The absorbing Markov chain for a focused packet delivery.}
\label{fig:absorbing_Markov}
\end{figure}

We denote by a probability vector $\mathbf{P}=(p_0,p_1,\cdots,p_{B-1})$ the steady state distribution of the corresponding relay queue of $y$ in a relay node $\mathbf{R}$, where each element $p_i$ denotes the probability that when $y$ enters the relay queue, there are $i$ packets already in this queue. Notice that the location of each node is stationary and ergodic with stationary distribution uniform on the network area, thus when $\mathbf{R}$ conducts the \textbf{R-D} transmission with probability $p_{rd}$ in a time slot, it will deliver a packet for each of the $n-2$ traffic flows with equal probability. Thus, if there are $i$ packets already in the relay queue of $y$, the expected time elapsed for $y$ to be delivered to its destination is $\displaystyle (i+1) \cdot \left( \frac{p_{rd}}{n-2} \right) ^{-1}$. Then we have 
\begin{align}
\overline{X}_R & \! = \! p_0 \cdot \frac{n-2}{p_{rd}} \!+\! 2p_1  \cdot \frac{n-2}{p_{rd}} \!+\! \cdots \!+\! B\cdot p_{B-1} \cdot \frac{n-2}{p_{rd}} \label{eq:X_R_1} \\
& \!=\!  \frac{n-2}{p_{rd}} \left \{ 1+p_1+2p_2 + \cdots+(B-1)p_{B-1} \right \} \label{eq:X_R_3} \\
& \!=\!  \frac{n-2}{p_{rd}} (1+ \overline{L}_{relay}^*),  \label{eq:X_R_4}
\end{align}
where (\ref{eq:X_R_3}) follows from the normalization property of a probability vector, and $\overline{L}_{relay}^*$ is the average queue length of a relay queue, under the condition that the relay buffer is not full.

We denote by $\mathbf{\Pi_L^*}=(\pi_{L_0}^*,\pi_{L_1}^*,\cdots,\pi_{L_{B-1}}^*)$ the limiting distribution of the level of a relay buffer, under the condition that the relay buffer is not full, then we have
\begin{equation}
\pi_{L_i}^*=\frac{\pi_{L_i}}{1-\pi_{L_B}}=\frac{\mathrm{C}_i \rho_s^i}{\sum_{k=0}^{B-1}{\mathrm{C}_k  \cdot \rho_s^k}},  \label{eq:conditional_distribution}
\end{equation}
and the corresponding conditional average number of packets occupying the relay buffer $\mathbb{E} \left \{L^* \right \}$ is given by
\begin{equation}
\mathbb{E} \left \{L^* \right \}= \sum_{i=0}^{B-1}{i\cdot \pi_{L_i}^*} = \frac{\sum_{i=0}^{B-1}{ i \mathrm{C}_i  \cdot \rho_s^i}}{\sum_{i=0}^{B-1}{\mathrm{C}_i  \cdot \rho_s^i}}=\Psi_{n,B,\lambda}.
\label{eq:conditional_level}
\end{equation}
Since these buffered packets are destined to each of the $n-2$ destinations with equal probability, then we have
\begin{equation}
\overline{L}_{relay}^* = \frac{\mathbb{E}\{ L^* \}}{n-2} \label{eq:conditional_relay_queue_length}.
\end{equation}

Substituting the results of (\ref{eq:X_R_4}), (\ref{eq:conditional_level}) and (\ref{eq:conditional_relay_queue_length}) into (\ref{eq:X_S}), the average transition times from the transient state $S$ to the absorbing state $D$ is determined as
\begin{equation}
\overline{X}_S=\frac{1+(n-2+\Psi_{n,B,\lambda})(1-p_o)}{\mu_s}.  \label{eq:X_S_1}
\end{equation}
Notice that $\displaystyle \mathbb{E}\{D\}=\overline{X}_S$, the result (\ref{eq:delivery_delay}) follows, and then the result (\ref{eq:end-to-end_delay}) follows from $\displaystyle \mathbb{E}\{T\}=\mathbb{E}\{Q\}+\mathbb{E}\{D\}$.

\end{IEEEproof}

\begin{remark}
Similar to the two-hop scenario, in the multi-hop MANETs the packet delivery process and occupancy behaviors of a relay buffer can be also modeled as an AMC and a QBD process respectively, so it is expected that our proposed theoretical framework for E2E delay modeling of two-hop MANETs can be also helpful for that of the multi-hop scenarios. It is notable, however, the state transition matrix of the QBD process will be different under the two scenarios, and also that with the multi-hop network scenarios there will be multiple transient states in the AMC.
\end{remark}

Based on Theorem~\ref{theorem:delay}, we can further extend our delay results to the buffer-unlimited scenario (i.e., $B \to \infty$), which is shown in the following corollary.

\begin{corollary} \label{corollary:infinite}
Considering the relay buffer size tends to infinity ($B \to \infty$), then $\mathbb{E}\{D\}$ and $\mathbb{E}\{T\}$ are determined as
\begin{align}
& \mathop{\mathbb{E}\{D\}}\limits_{B \to \infty}= \frac{1}{p_{sd}+p_{sr}}+ \frac{n-2}{p_{sd}+p_{sr}-\lambda}, 
\label{eq:delivery_delay_infinite} \\
& \mathop{\mathbb{E}\{T\}}\limits_{B \to \infty} = \frac{n-1-\lambda}{p_{sd}+p_{sr}-\lambda}. 
\label{eq:end-to-end_delay_infinite}
\end{align}

\end{corollary}  

\begin{IEEEproof}
See Appendix~\ref{appendix:infinite} for the proof.
\end{IEEEproof}

\begin{remark}
Notice that when $B\to \infty$, the results of Lemma~\ref{lemma:throughput_capacity} and Corollary~\ref{corollary:infinite} are coincident with the capacity and delay derived in \cite{Neely_IT05}, where the relay buffer size is assumed to be infinite.
\end{remark}

\section{Case Studies} \label{section:case_studies}

\begin{figure}[!t]
\centering
\includegraphics[width=3.2in]{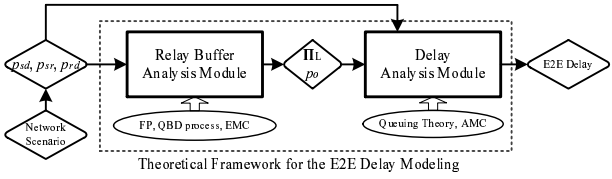} \caption{Illustration of the application of our theoretical framework.}
\label{fig:framework}
\end{figure}

In this section, we conduct case studies to illustrate the application of our theoretical framework for the E2E delay modeling in buffer-limited MANETs. As illustrated in Fig.~\ref{fig:framework}, for a given network scenario, the corresponding $p_{sd}$, $p_{sr}$ and $p_{rd}$ should be determined first, then with the inputs of these probabilities, by sequentially executing the relay buffer analysis module and delay analysis module, this framework finally returns the delay results. The details of the application of our framework under specific network scenarios are shown in the following subsections.

\subsection{Cell-partitioned MANET with LS-MAC}

We first consider a cell-partitioned MANET with local scheduling based MAC protocol (LS-MAC) \cite{Neely_IT05}. The whole network area is partitioned into $m \times m$ non-overlapping cells of equal size. In a time slot, each cell can support only one pair of nodes for packet transmission, concurrent transmissions in different cells will not interference with each other, and nodes within different cells cannot communicate. At the beginning of each time slot, all nodes in a cell contends for the wireless channel access using a DCF-style mechanism \cite{Bianchi_JSEC00}. In addition to these network settings, the MANET also meets the set of assumptions described in Section~\ref{subsection:assumption}.
 
With the detailed information of network settings, we then determine the corresponding probabilities $p_{sd}$, $p_{sr}$ and $p_{rd}$, provided in the following lemma.

\begin{lemma} \label{lemma:probabilities_LS}
For the concerned cell-partitioned MANET with LS-MAC, the probabilities $p_{sd}$, $p_{sr}$ and $p_{rd}$ are given by
\begin{equation}
p_{sd} \!=\! \frac{m^2}{n}\!-\!\frac{m^2-1}{n-1}\!+\!(\frac{m^2-1}{n-1}\!-\!\frac{m^2-1}{n})(1\!-\!\frac{1}{m^2})^{n-1},    \label{eq:p_sd_LS} 
\end{equation}
\begin{align}
p_{sr}  & \!= \! p_{rd} \nonumber \\
        & \!=\!  \frac{1}{2}\left \{ \frac{m^2-1}{n-1} \! - \! \frac{m^2}{n-1} (1\!-\!\frac{1}{m^2})^n \! - \!(1\!-\! \frac{1}{m^2})^{n-1} \right \}. \label{eq:p_sr_LS}
\end{align}

\end{lemma}

\begin{IEEEproof}
See Appendix~\ref{appendix:basic_probabilities} for the proof.
\end{IEEEproof}

Given the number of nodes $n$ and relay buffer size $B$, substituting formulas~(\ref{eq:p_sd_LS}) and (\ref{eq:p_sr_LS}) into formula~(\ref{eq:throughput_capacity}), we first determine the throughput capacity $\mu$ of such a MANET. Then with any packet generating rate $\lambda<\mu$, we substitute formulas~(\ref{eq:p_sd_LS}) and (\ref{eq:p_sr_LS}) into equation~(\ref{eq:ROP}) to determine the corresponding ROP $p_o$, and $\mu_s$ can be further determined by formula~(\ref{eq:mu_s}). Substituting $\lambda$, $p_o$ and $\mu_s$ into formulas~(\ref{eq:queuing_delay}), (\ref{eq:delivery_delay}) and (\ref{eq:end-to-end_delay}), we finally obtain the results of queuing delay, delivery delay and E2E delay respectively for the concerned buffer-limited MANET.

\subsection{Cell-partitioned MANET with Power Control and EC-MAC} 

\begin{figure}
    \centering{        
    \subfigure[Transmission range of a node.]	
		{\includegraphics[width=1.65in]{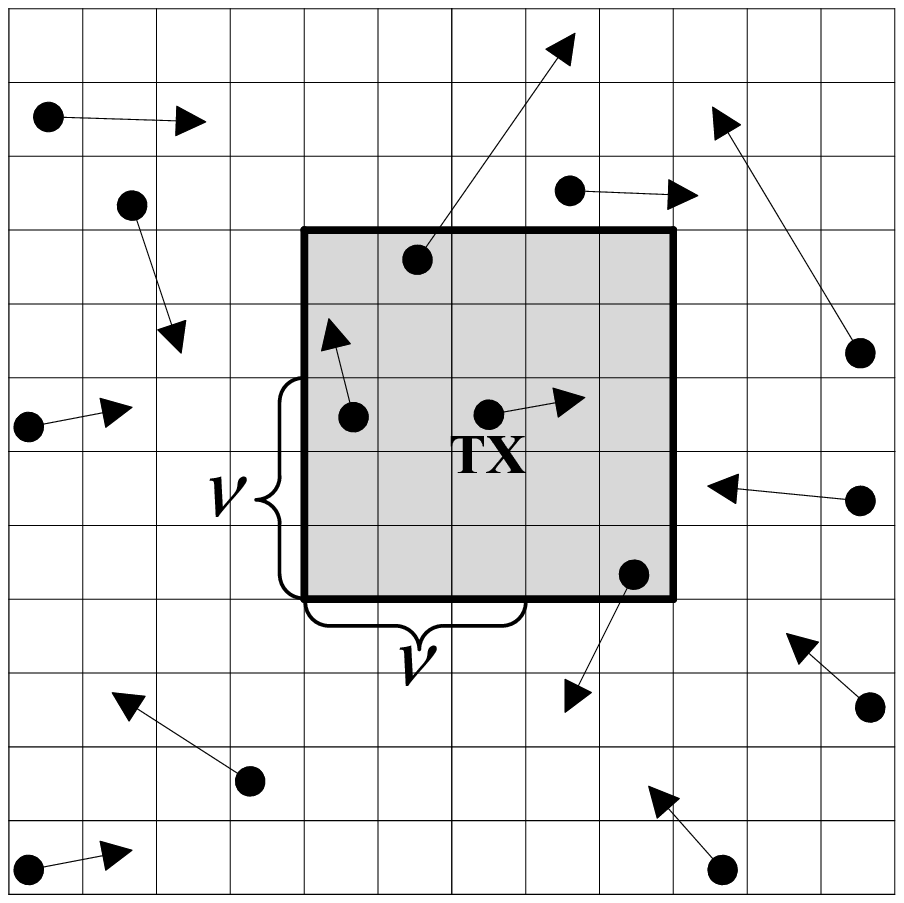} \label{fig:power_control} }
		\hfill
    \subfigure[Illustration of EC-MAC scheduling.]		
		{\includegraphics[width=1.65in]{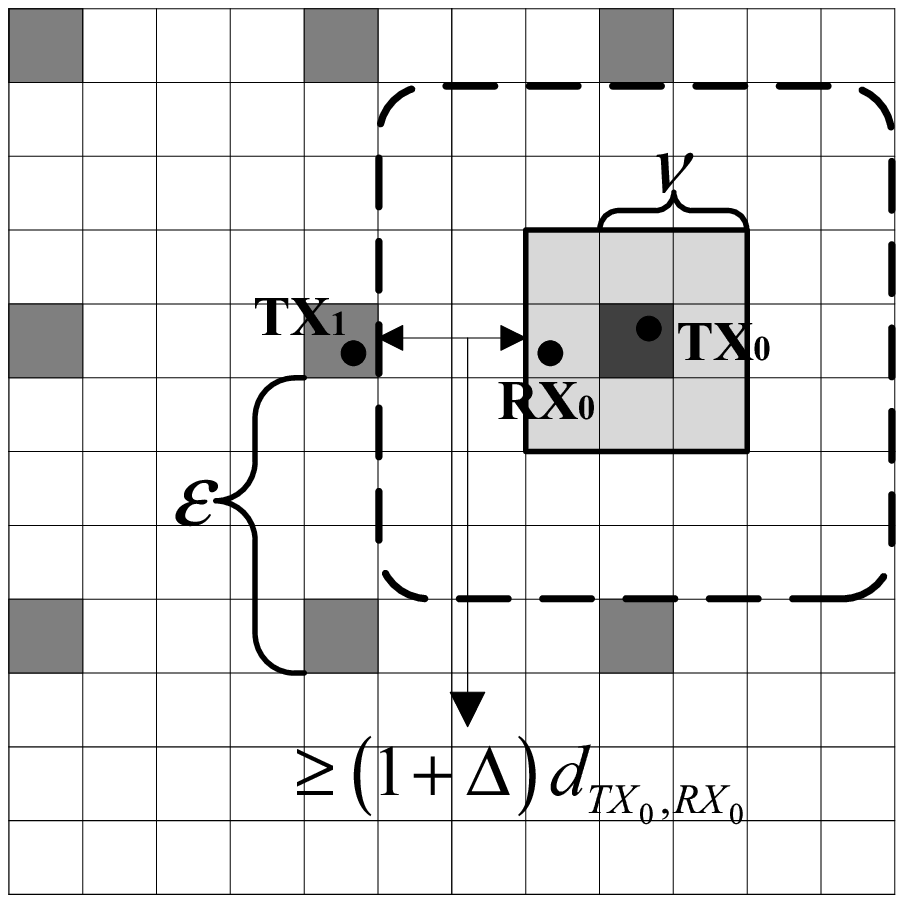}\label{fig:EC_MAC}}
    \caption{A cell-partitioned MANET with power control and EC-MAC}
		}
\end{figure}

We then consider a more general cell-partitioned MANET which applies the power control and Equivalent-Class based MAC protocol (EC-MAC) \cite{Liu_WCNC12, Kulkarni_IT04,Gao_IEICE13}. As shown in Fig.~\ref{fig:power_control}, the transmission range of a transmitter $TX$ covers a set of cells which have a horizontal and vertical distance of no more than $\nu-1$ cells away from its own cell. Meanwhile, as illustrated in Fig.~\ref{fig:EC_MAC}, all cells are divided into different ECs, where any two cells in the same EC have a horizontal and vertical distance of some multiple of $\varepsilon$ cells. Thus, the MANET contains in total $\varepsilon ^2$ ECs and ECs are activated alternatively as time evolves. Suppose that at time slot $t$, a transmitter $TX_0$ in an active cell will transmit a packet to its receiver $RX_0$, in order to ensure the transmission successful, according to the Protocol Model \cite{Gupta_IT00} it should satisfy that
\begin{equation}
d_{TX_1,RX_0} \geq (1+\Delta) d_{TX_0,RX_0}, 
\end{equation}
where $TX_1$ denotes a concurrent transmitter in any one of the other active cells, $d_{i,j}$ denotes the distance between nodes $i$ and $j$, and $\Delta$ is a guard factor. Thus we have
\begin{equation}
\varepsilon-\nu \geq (1+\Delta) \sqrt{2} \nu , \label{eq:epsilon}
\end{equation}
and $\varepsilon$ should be set as
\begin{equation}
\varepsilon=\min \{\lceil (1+\Delta) \sqrt{2} \nu+\nu \rceil,m \}. \label{eq:set_epsilon}
\end{equation}

Regarding the corresponding probabilities $p_{sd}$, $p_{sr}$ and $p_{rd}$ of this type of MANETs, we have the following lemma.

\begin{lemma} \label{lemma:probabilities_EC}
For the concerned MANET with power control and EC-MAC, the probabilities $p_{sd}$, $p_{sr}$ and $p_{rd}$ are given by
\begin{equation}
p_{sd} \!=\! \frac{1}{\varepsilon^2} \left \{ \frac{\Gamma\!-\!\frac{m^2}{n}}{n-1} \! + \! \frac{m^2\!-\!1\!-\!(\Gamma\!-\!1)n}{n(n-1)} (1\!-\! \frac{1}{m^2}) ^{n-1} \right \},  \label{eq:p_sd_EC} 
\end{equation}
\begin{align}
p_{sr} & \!=\!p_{rd} \nonumber \\
       & \!=\! \frac{1}{2 \varepsilon^2} \left \{ \frac{m^2\!-\!\Gamma}{n\!-\!1} (1\!-\!(1\!-\! \frac{1}{m^2})^{n\!-\!1}) \!-\! (1\!-\! \frac{\Gamma}{m^2})^{n\!-\!1}     \right \}, \label{eq:p_sr_EC}
\end{align}
where $\Gamma=(2\nu-1)^2$.
\end{lemma}

\begin{IEEEproof}
See Appendix~\ref{appendix:basic_probabilities} for the proof.
\end{IEEEproof}

By applying the same operations of our theoretical framework as the previous subsection, we can obtain the delay results for the concerned MANETs.

\subsection{Other Network Scenarios}
Notice that to apply our theoretical framework for the E2E delay modeling, it only needs to determine the inputs of the framework, i.e., the probabilities $p_{sd}$, $p_{sr}$ and $p_{rd}$. Thus, this framework also has the great potential to be applied to many other network scenarios. For example, for the MANETs where the 2HR routing scheme is administered by cell \cite{Neely_IT05} (not by each node in our case), and the ratio between \textbf{S-R} and \textbf{R-D} transmission can be changed \cite{Gao_IEICE13} (we fix the ratio as $0.5$), these probabilities can be determined. Recently, Chen~\emph{et al.} \cite{Chen_ICCC13} has reported that how to compute these probabilities for a MANET under the continuous network model and the Aloha MAC protocol, then with these probabilities as the inputs of our framework, the corresponding delay analysis under the practical buffer constraint can be conducted.

\section{Simulation Results} \label{section:simulation}
In this section, we first conduct simulations to validate our theoretical framework for the E2E delay modeling in buffer-limited MANETs, then provide discussions about the impacts of network parameters on delay performance.

\subsection{Simulation Settings}

For the validation of our theoretical framework and delay results, a specific C++ simulator was developed to simulate the packet generating, queuing and delivery processes in a cell-partitioned MANET \cite{C++}, where the network settings, including relay buffer size $B$, number of nodes $n$, partition parameter $m$, packet generating rate $\lambda$ and the mobility model can be flexibly adjusted to simulate the network performance under various scenarios. For the network scenario with power control and EC-MAC, we set $\nu=1$ and $\Delta=1$ \cite{ns-2}. The duration of each task of simulation is set to be $2\times 10^8$ time slots, and we only collect data from the last $80\%$ of the time slots in each task (the system will be in the steady state with high probability), to ensure the accuracy of simulated results.

\subsection{Validation}

\begin{figure}
    \centering
    {
    \subfigure[ROP under LS-MAC.]
    {\includegraphics[width=3.0in]{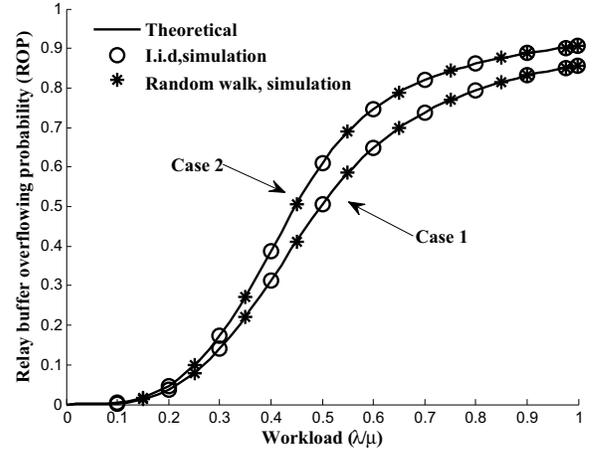} \label{subfig:ROP_LS}}
   	\hfill
    \subfigure[ROP under power control and EC-MAC.]
    {\includegraphics[width=3.0in]{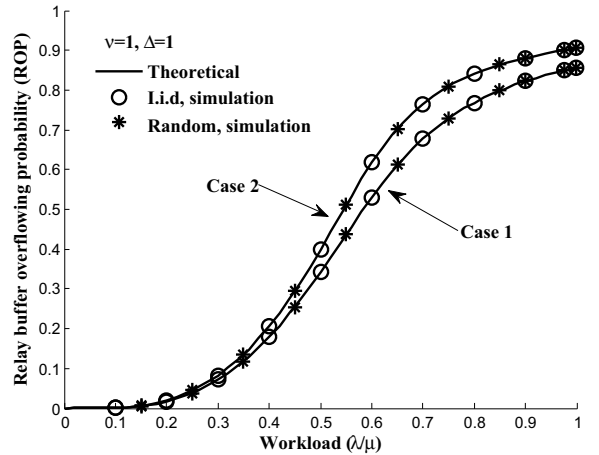} \label{subfig:ROP_EC}}
    }
    \caption{Theoretical and simulated ROP performance. Case 1: $n=32, m=4, B=5$. Case 2: $n=50, m=5, B=5$.}
    \label{fig:ROP_simulation}
\end{figure}

First, we provide plots of the theoretical and simulated ROP performance under two network scenarios in Fig.~\ref{fig:ROP_simulation}, and for each scenario we consider two cases (case 1: $n=32, m=4, B=5$, and case 2: $n=50, m=5, B=5$) and two mobility models (the i.i.d mobility model and the random walk model). The workload is defined as $\displaystyle \lambda / \mu$. We can see from Fig.~\ref{fig:ROP_simulation} that the simulation results match nicely with the theoretical ones for all the cases, which indicates that our framework is highly efficient in depicting the occupancy behaviors of the relay buffer in
buffer-limited MANETs. 

\begin{figure}
    \centering
    {
    \subfigure[End-to-end delay under LS-MAC.]
    {\includegraphics[width=3.0in]{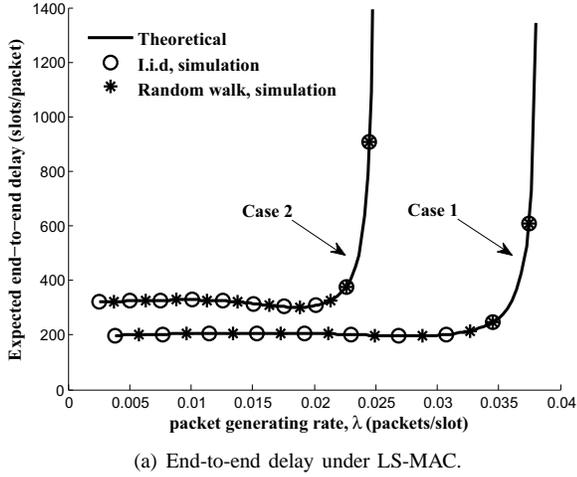} \label{subfig:delay_LS}}
   	\hfill
    \subfigure[End-to-end delay under power control and EC-MAC.]
    {\includegraphics[width=3.0in]{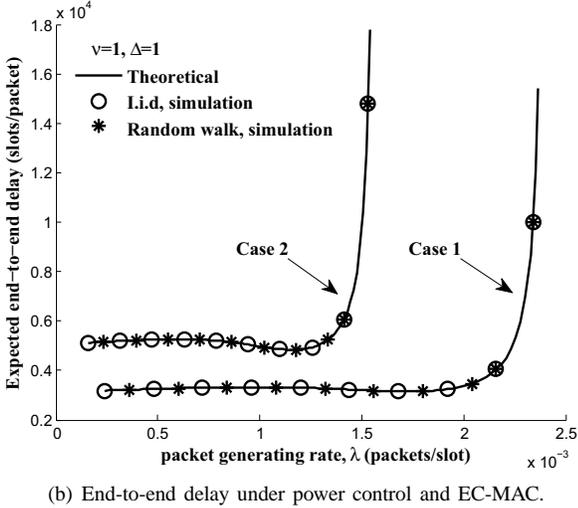} \label{subfig:delay_EC}}
    }
    \caption{Theoretical and simulated end-to-end delay performance. Case 1: $n=32, m=4, B=5$. Case 2: $n=50, m=5, B=5$.}
    \label{fig:delay}
\end{figure}

Then, with the same network settings, we provide plots of the theoretical and simulated E2E delay results in Fig.~\ref{fig:delay}. It is observed from Fig.~\ref{fig:delay} that all the simulation results can match the corresponding theoretical curves very nicely, indicating that: 1) our theoretical framework is highly efficient for the E2E delay modeling in buffer-limited MANETs; 2) the framework is very general since it can be applied to various network scenarios. Another observation of Fig.~\ref{fig:delay} is that the packet E2E delay increases sharply as the packet generating rate $\lambda$ approaches a specific value (e.g., under LS-MAC and case 1, the value is around 0.038), which serves as an intuitive impression of its corresponding throughput capacity $\mu$. 

\begin{figure}[!t]
\centering
\includegraphics[width=3.0in]{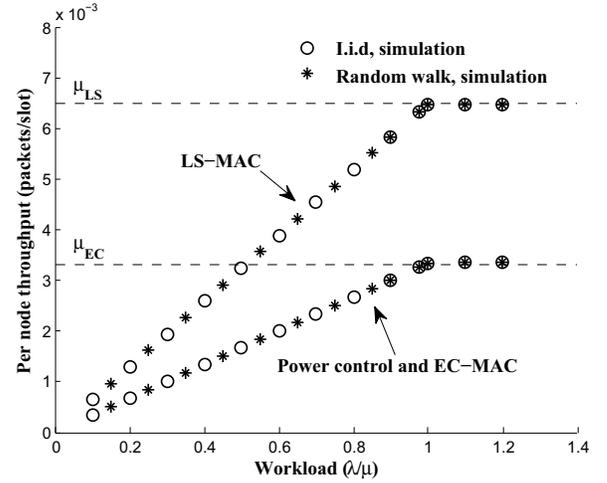} 
\caption{Per node achievable throughput. LS-MAC: $n=200, m=10, B=5$. Power control and EC-MAC: $n=32, m=4, B=10$.}
\label{fig:throughput}
\end{figure}

To further validate our framework on throughput capacity, Fig.~\ref{fig:throughput} summarizes the simulation results on the achievable per node throughput, where two network scenarios with different throughput capacity (LS-MAC: $n=200, m=10, B=5, \mu_{LS}=6.5\times 10^{-3}$) and (power control and EC-MAC: $n=32, m=4, B=10, \mu_{EC}=3.3\times 10^{-3}$) are presented. We can see that for each network scenario there, the per node throughput first monotonously increases before the workload reaches $1$, and then remains a constant which is just the corresponding throughput capacity when the workload exceeds $1$. Thus, it indicates that the proposed framework is also efficient in depicting the per node throughput capacity behavior of the buffer-limited MANET.

\subsection{Performance Discussion}

With the help of our theoretical framework for the E2E delay modeling, we explore how the network parameters affect the the delay performance of a buffer-limited MANET. Without loss of generality, we consider here a cell-partitioned MANET with LS-MAC.

\begin{figure}[!t]
\centering
\includegraphics[width=3.0in]{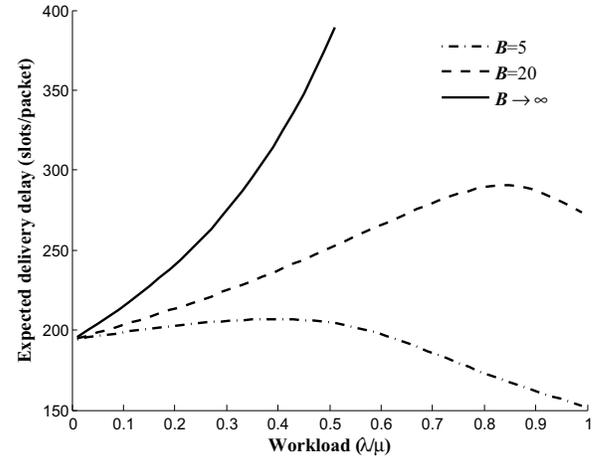} 
\caption{Delivery delay vs. workload ($\lambda/\mu$) under different settings of relay buffer size. $n=32, m=4$.}
\label{fig:delivery_delay}
\end{figure}

We first summarize in Fig.~\ref{fig:delivery_delay} that how the expected delivery delay $\mathbb{E}\{D\}$ varies with the workload. A very interesting observation is that under the buffer-limited scenarios ($B=5$ and $B=20$), as workload increases, $\mathbb{E}\{D\}$ first increases to a maximum and then decreases. This is due to the reason that the effects of workload on $\mathbb{E}\{D\}$ are two folds. On one hand, a larger workload will lead to a longer relay queue length, which further leads to a higher delay in the relay queue; on the other hand, a larger workload will lead to a higher ROP, which further leads to a lower probability that a packet to be delivered by a two-hop way, such that $\mathbb{E}\{D\}$ decreases. Since the latter effect, the delivery delay under a small relay buffer is lower than that under a large one. 

\begin{figure}[!t]
\centering
\includegraphics[width=3.0in]{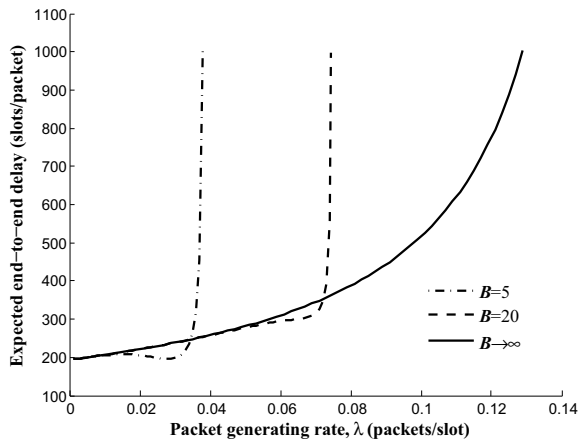} 
\caption{End-to-end delay vs. packet generating rate $\lambda$ under different settings of relay buffer size. $n=32, m=4$.}
\label{fig:end-to-end_delay_vs_lambda}
\end{figure}

\begin{figure}[!t]
\centering
\includegraphics[width=3.0in]{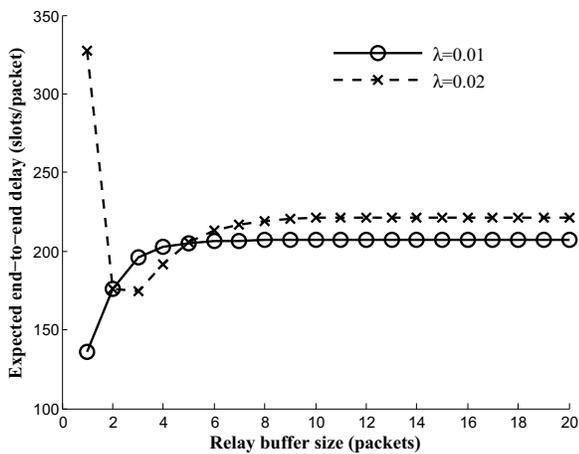} 
\caption{End-to-end delay vs. relay buffer size. $n=32, m=4$.}
\label{fig:end-to-end_delay_vs_buffer}
\end{figure}

Fig.~\ref{fig:end-to-end_delay_vs_lambda} shows the relationship between the expected E2E delay $\mathbb{E}\{T\}$ and packet generating rate $\lambda$. We can see that under buffer-limited scenarios, as $\lambda$ increases, $\mathbb{E}\{T\}$ doesn't increase all the time because the delivery delay will decrease when $\lambda$ exceeds a specific value, however when $\lambda$ approaches the corresponding throughput capacity, $\mathbb{E}\{T\}$ increases sharply because the queuing delay tends to infinity. It also can be seen that when $\lambda$ is small, $\mathbb{E}\{T\}$ under $B=5$ is smaller than that under $B=20$, since both of the queuing delay under two settings are small, but a small relay buffer can lead to a small delivery delay. However, with $\lambda$ getting larger and larger, $\mathbb{E}\{T\}$ under $B=5$ finally exceeds that under $B=20$, and tends to infinity earlier. It indicates that increasing the relay buffer size can ensure the E2E delay limited for a larger region of packet generating rate.  

We illustrate in Fig.~\ref{fig:end-to-end_delay_vs_buffer} how $\mathbb{E}\{T\}$ varies $B$ under the settings of $(n=32, m=4, \lambda=\{0.01, 0.02\})$. According to formula~(\ref{eq:throughput_capacity}), $\mu=0.0227$ when $B=1$, and $\mu$ increases as $B$ increases. Thus, for $\lambda=0.01$ which is much smaller than $0.0227$, $\mathbb{E}\{T\}$ increases as $B$ increases and finally tends to a constant $206.92$ which can be determined by formula~(\ref{eq:end-to-end_delay_infinite}). While for $\lambda=0.02$ which is very close to the $\mu$ under $B=1$, $\mathbb{E}\{T\}$ under $B=1$ is very large. With $B$ increasing, the corresponding $\mu$ increases, leading to the $\mathbb{E}\{T\}$ first decreases, then increases and finally tends to a constant $221.65$.

\begin{figure}[!t]
\centering
\includegraphics[width=3.0in]{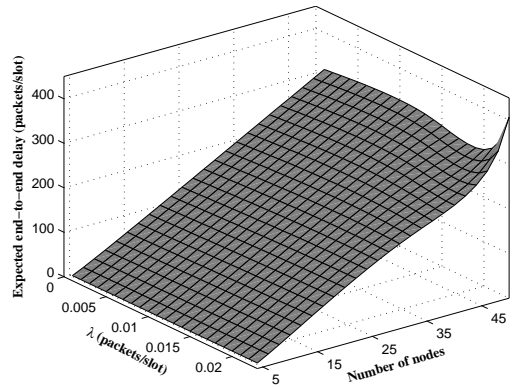} 
\caption{End-to-end delay vs. packet generating rate $\lambda$ and number of nodes $n$. $B=5$.}
\label{fig:delay_lambda_node} 
\end{figure}

\begin{figure}[!t]
\centering
\includegraphics[width=3.0in]{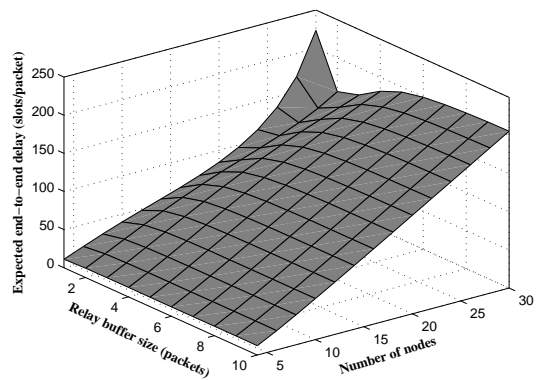} 
\caption{End-to-end delay vs. relay buffer size $B$ and number of nodes $n$. $\lambda=0.02$.}
\label{fig:delay_buffer_node}
\end{figure}

We further illustrate in two 3D figures (Fig.~\ref{fig:delay_lambda_node} and Fig.~\ref{fig:delay_buffer_node}) that how $\mathbb{E}\{T\}$ is influenced by $\{n, \lambda\}$ and $\{n,B\}$, respectively (the ratio of $n$ to the number of cells keeps as $2$). We can see that the variations of $\mathbb{E}\{T\}$ with  $n$ are complicated, but in general $\mathbb{E}\{T\}$ increases as $n$ increases. A more careful observation is that when $n$ increases, $\mathbb{E}\{T\}$ first increases almost linearly when $\lambda$ is much smaller than $\mu$, then increases quickly when $\lambda$ approaches $\mu$. For example, these behaviors can be found in Fig.~\ref{fig:delay_lambda_node} under $\lambda=0.23$ and in Fig.~\ref{fig:delay_buffer_node} under $B=1$.

\section{Conclusion} \label{section:conclusion}
This paper represents a significant step towards the exact end-to-end delay modeling of practical buffer-limited MANETs. With the help of the theories of Fixed-Point, QBD process, EMC and AMC, a novel theoretical framework has been developed to efficiently depict the highly dynamics in such networks. This framework is general in the sense that it can be applied to the MANETs with any mobility model that leads to the uniform distribution of the locations of nodes, and any MAC protocol as long as the probabilities $p_{sd}$, $p_{sr}$ and $p_{rd}$ there can be determined. Also, it is expected the framework can shed light on the E2E delay modeling for multi-hop MANETs. Extensive simulations have been conducted to validate the efficiency and applicability of our framework, and some interesting theoretical findings about the impacts of network parameters on delay performance have been discussed. 

Notice that our E2E delay modeling for buffer-limited MANETs is based on the 2HR routing and relay buffer constraint, so one of the future research directions is to extend our study to the E2E delay modeling for MANETs with multi-hop routing schemes and more practical buffer constraint on both source buffer and relay buffer. Another appealing direction is to explore the delay modeling for the concerned MANETs with the consideration of more practical network settings (such as the wireless channel fading) and apply the well-known NS-2 network simulator for model validation.


\begin{appendices}

\section{Proof of Lemma~\ref{lemma:EMC_transition}}\label{appendix:EMC_transition}

We denote by $\lambda_R$ the packet arrival rate at the relay buffer in $\mathbf{S}$. Due to the ergodic and uniform properties of node mobility, for node $\mathbf{S}$, each of the remaining $n-2$ nodes (except $\mathbf{D}$) is likely to serve as its relay with equal probability. Due to the symmetry of nodes and traffic flows, for $\mathbf{S}$ serving as a relay, all the other $n-2$ nodes are likely to forward packets to its relay buffer. Notice that the packet departure process of source queue is a Bernoulli process with rate $\lambda$, and from (\ref{eq:mu_s}) we can see the ratio of \textbf{S-R} transmission to the whole packet departure is $\frac{p_{sr}\left ( 1-p_o(\lambda)\right )}{\mu_s(\lambda)}$, then we have
\begin{align}
\lambda_R &= (n-2)\lambda \cdot \frac{p_{sr}\left(1-p_o(\lambda)\right)}{\mu_{s}(\lambda)} \big/ (n-2) \nonumber \\
          & = \rho_s(\lambda) p_{sr} \left( 1-p_o(\lambda) \right).
\label{eq:lambda_R_1}
\end{align} 

Regarding the transition probability from state $(i,j)$ to the states in its adjacent upper level $L_{i+1}$, we have the following equation
\begin{equation}
p_{(i,j),L_{i+1}} \cdot \left( 1-p_o(\lambda) \right) + 0 \cdot p_o(\lambda)=\lambda_R.  
\label{eq:lambda_R}
\end{equation}
Combining (\ref{eq:lambda_R}) with (\ref{eq:lambda_R_1}) we have
\begin{equation}
p_{(i,j),L_{i+1}}=\rho_s(\lambda)p_{sr}.  
\label{eq:p_ij_i+1}
\end{equation}
Substituting (\ref{eq:p_ij_i+1}) into (\ref{eq:phase_average}) we have
\begin{equation}
p_L^{(i,i+1)}= \rho_s(\lambda)p_{sr}.  \nonumber
\end{equation}

Regarding the transition probability from state $(i,j)$ to the states in its adjacent lower level $L_{i-1}$, when $\mathbf{S}$ conducts a \textbf{R-D} transmission with probability $p_{rd}$, due to the ergodic and uniform features of node mobility, it will choose one of the other $n-2$ nodes as its receiver with equal probability. Then we have 
\begin{equation}
p_{(i,j),L_{i-1}}=p_{rd}  \cdot \frac{j}{n-2}.
\label{eq:p_ij_i-1}
\end{equation}

To determine the conditional probability $P_{j|L_i}$, we utilize the following Occupancy technique \cite{Stark_BOOK02}. Considering the relay buffer on level $i$, where each of these $i$ buffered packets may be destined for any one of the other $n-2$ nodes (except $\mathbf{S}$ and $\mathbf{D}$), the number of all possible cases $N_{L_i}$ is determined as
\begin{equation}
N_{L_i}=\binom{n-3+i}{i}.
\label{eq:events_all}
\end{equation}
Considering the condition that these $i$ packets are destined for only $j$ different nodes, then the number of possible cases $N_{(i,j)}$ is determined as
\begin{equation}
N_{(i,j)}=\binom{n-2}{j}\cdot \binom{(j-1)+(i-j)}{i-j}.
\label{eq:events_part}
\end{equation}
Due to the ergodic and uniform features of node mobility, each of these cases occurs with equal probability. According to the \emph{Classical Probability}, $P_{j|L_i}$ is then determined as 
\begin{equation}
P_{j|L_i}=\frac{N_{(i,j)}}{N_{L_i}}
=\frac{\binom{n-2}{j}\cdot\binom{i-1}{j-1}}{\binom{n-3+i}{i}}.
\label{eq:conditional_pro}
\end{equation}
It can be easily verified that $\sum \limits_{j\leq i} {P_{j|L_i}}=1$. Combining (\ref{eq:p_ij_i-1}), (\ref{eq:conditional_pro}) and (\ref{eq:phase_average}) we have
\begin{align}
p_L^{(i,i-1)} & =\sum_{j=1}^{i} \left\{ \frac{ \binom{n-2}{j} \cdot \binom{i-1}{j-1}} {\binom{n-3+i}{i}}\cdot\ p_{rd} \frac{j}{n-2}\right\} \nonumber \\
              & = \frac{p_{rd}}{\binom{n-3+i}{i}} \cdot \sum_{j=0}^{i-1}\left\{ \binom{n-3}{j} \cdot \binom{i-1}{j}\right\} \nonumber \\
              & = p_{rd} \cdot \frac{\binom{n-4+i}{i-1}}{\binom{n-3+i}{i}} = \frac{i}{n-3+i} \cdot p_{rd}.
\end{align}

\section{Proof of Lemma~\ref{lemma:throughput_capacity}}\label{appendix:throughput_capacity}
Since the relay buffer is limited, it is always stable. Thus, we only need to consider the stability of source queue. The source queue is a Bernoulli/Bernoulli queue with packet generating rate $\lambda$ and a corresponding service rate $\mu_s(\lambda)$. Based on the queuing theory, the source queue is stable (resp. unstable) when $\lambda < \mu_s(\lambda)$ (resp. $\lambda \geq \mu_s(\lambda)$).

As $\lambda$ increases, $p_o(\lambda)$ increases since the network is more congested, and $\mu_s(\lambda)$ decreases according to formula (\ref{eq:mu_s}). Thus, from the definition of throughput capacity, $\mu$ should satisfy that
\begin{equation}
\mu=\lambda^*=\mu_s(\lambda^*). \label{eq:mu}
\end{equation} 
It is notable that when $\lambda$ approaches $\lambda^*$, $\displaystyle \lim \limits_{\lambda \to \lambda^*}\rho_s(\lambda)=1$, and from (\ref{eq:ROP}) we have
\begin{equation}
p_o(\lambda^*)=\frac{\mathrm{C}_B} {\sum_{k=0}^B{\mathrm{C}_i }}=\frac{n-2}{n-2+B}. \label{eq:ROP_mu}
\end{equation}
Then $\mu$ is determined by substituting (\ref{eq:ROP_mu}) into (\ref{eq:mu_s}).


\section{Proof of Corollary~\ref{corollary:infinite}} \label{appendix:infinite}

We denote by $F(\rho_s)$, $G(\rho_s)$ the sums of infinite series $ \sum _{i \geq 0}  \mathrm{C}_i \rho_s^i$ and $\sum_{i \geq 0} i \mathrm{C}_i \rho_s^i$, respectively. Notice that $F(\rho_s)$ is the Taylor series expansion from $(1-\rho_s)^{2-n}$, then we have
\begin{align}
& F(\rho_s)=\frac{1}{(1-\rho_s)^{n-2}}, \\
& G(\rho_s)=\rho_s \cdot F'(\rho_s)= (n-2)\frac{\rho_s}{(1-\rho_s)^{n-1}}.
\end{align}
Further we have
\begin{equation}
\Psi_{n,\infty,\lambda}=\frac{G(\rho_s)}{F(\rho_s)}=(n-2)\frac{\rho_s}{1-\rho_s}, \label{eq:level_infinite} 
\end{equation}
and
\begin{align}
\mathop{p_o}\limits_{B \to \infty} & = \lim \limits_{B \to \infty} \mathrm{C}_B \cdot \rho_s^B \cdot (1-\rho_s)^{n-2} \\
& \leq \lim \limits_{B \to \infty} (B+n)^n \rho_s^B \leq \lim \limits_{B \to \infty} 2^n B^n \rho_s^B \\
& = \lim \limits_{B \to \infty} \frac{n! \rho_s^B}{(-\ln \rho_s)^n}=0, \label{eq:po_infinite}
\end{align}
where (\ref{eq:po_infinite}) is obtained by utilizing the L'H{\^o}pital's rule recursively.

Substituting (\ref{eq:level_infinite}) and (\ref{eq:po_infinite}) into Theorem~\ref{theorem:delay}, we can obtain (\ref{eq:delivery_delay_infinite}) and (\ref{eq:end-to-end_delay_infinite}) directly.

\section{Proofs of Lemma~\ref{lemma:probabilities_LS} and Lemma~\ref{lemma:probabilities_EC}} \label{appendix:basic_probabilities}

For a cell-partitioned MANET with LS-MAC, the event that node $\mathbf{S}$ conducts a \textbf{S-D} (resp. \textbf{S-R} or \textbf{R-D}) transmission in a time slot can be divided into the following sub-events: (1) $\mathbf{D}$ is (resp. is not) in the same cell with $\mathbf{S}$; (2) other $k$ out of $n-2$ nodes are in the same cell with $\mathbf{S}$, while the remaining $n-2-k$ nodes are not in this cell; (3) $\mathbf{S}$ contends for the wireless channel access successfully. Thus we have
\begin{align*}
p_{sd} & =\sum_{k=0}^{n-2}{\binom{n-2}{k} (\frac{1}{m^2})^{k+1} (1-\frac{1}{m^2})^{n-2-k}  \cdot \frac{1}{k+2}}   \\ 
       & =\sum_{k=0}^{n-2}{\binom{n-1}{k+1} (\frac{1}{m^2})^{k+1} (1-\frac{1}{m^2})^{n-2-k}  \cdot \frac{1}{k+2}} \\ 
			 & -\sum_{k=0}^{n-3}{\binom{n-2}{k+1} (\frac{1}{m^2})^{k+1} (1-\frac{1}{m^2})^{n-2-k}  \cdot \frac{1}{k+2}} \\
			 & =\frac{m^2}{n}\left \{ 1-(1-\frac{1}{m^2})^n \right \}-(1-\frac{1}{m^2})^{n-1}                           \\ 
			 & -\frac{m^2-1}{n-1}\left \{ 1-(1-\frac{1}{m^2})^{n-1} \right \} +(1-\frac{1}{m^2})^{n-1}                  \\ 
			 & =\frac{m^2}{n}-\frac{m^2-1}{n-1}+(\frac{m^2-1}{n-1}-\frac{m^2-1}{n})(1-\frac{1}{m^2})^{n-1}, \nonumber
\end{align*}
and
\begin{align*}
p_{sr} & =p_{rd}  \\
       & =\frac{1}{2}\sum_{k=1}^{n-2}{\binom{n-2}{k} (\frac{1}{m^2})^k (1-\frac{1}{m^2})^{n-1-k} \cdot \frac{1}{k+1}}  \\
       & =\frac{1}{2}\left \{ \frac{m^2-1}{n-1}-\frac{m^2}{n-1} (1-\frac{1}{m^2})^n-(1-\frac{1}{m^2})^{n-1} \right \}  
\end{align*}

For a cell-partitioned MANET with power control and EC-MAC, by applying the similar approach and algebraic operations we have
\begin{align*}
p_{sd} & = \frac{1}{\varepsilon ^2} \left \{ \sum_{k=0}^{n-2} {\binom{n-2}{k}(\frac{1}{m^2})^{k+1}(1-\frac{1}{m^2})^{n-2-k} \cdot \frac{1}{k+2}} \right.  \\
       & + \left. \sum_{k=0}^{n-2} {\binom{n-2}{k} (\frac{1}{m^2})^{k+1} (1-\frac{1}{m^2})^{n-2-k}  \cdot \frac{4v^2-4v}{k+1} } \right \}  \\
			 & = \frac{1}{\varepsilon^2} \left \{ \frac{\Gamma-\frac{m^2}{n}}{n-1} + \frac{m^2-1-(\Gamma-1)n}{n(n-1)} (1-\frac{1}{m^2}) ^{n-1}  \right \}, 
\end{align*}
and
\begin{align*} 
p_{sr} &=p_{rd}  \\
       &= \frac{1}{2 \varepsilon^2} \frac{m^2-\Gamma}{m^2} \cdot \\
       &  \left \{ \sum_{k=1}^{n-2}{ \binom{n-2}{k} (\frac{1}{m^2})^k (1-\frac{1}{m^2})^{n-2-k} \cdot \frac{1}{k+1}}\right.  \\
       & + \left. \sum_{k=1}^{n-2} { \binom{n-2}{k} (\frac{\Gamma-1}{m^2})^k (\frac{m^2-\Gamma}{m^2})^{n-2-k} } \right\}  \\
			 & = \frac{1}{2 \varepsilon^2} \left \{ \frac{m^2-\Gamma}{n-1} (1-(1-\frac{1}{m^2})^{n-1}) - (1-\frac{\Gamma}{m^2})^{n-1}     \right \} .
\end{align*}

\end{appendices}


\end{document}